# Protection or Peril of Following the Crowd in a Pandemic-Concurrent Flood Evacuation

**Elisa Borowski**

Department of Civil and Environmental Engineering, Northwestern University, Technological Institute, 2145 Sheridan Road, Evanston, IL 60208, Email: elisaborowski2022@u.northwestern.edu

**Amanda Stathopoulos, Ph.D.** (Corresponding Author)

Department of Civil and Environmental Engineering, Northwestern University, Technological Institute, 2145 Sheridan Road, Evanston, IL 60208, Email: a-stathopoulos@northwestern.edu

## ABSTRACT

The decisions of whether and how to evacuate during a climate disaster are influenced by a wide range of factors, including sociodemographics, emergency messaging, and social influence. Further complexity is introduced when multiple hazards occur simultaneously, such as a flood evacuation taking place amid a viral pandemic that requires physical distancing. Such multi-hazard events can necessitate a nuanced navigation of competing decision-making strategies wherein a desire to follow peers is weighed against contagion risks. To better understand these nuances, we distributed an online survey during a pandemic surge in July 2020 to 600 individuals in three midwestern and three southern states in the United States with high risk of flooding. In this paper, we estimate a random parameter logit model in both preference space and willingness-to-pay space. Our results show that the directionality and magnitude of the influence of peers' choices of whether and how to evacuate vary widely across respondents. Overall, the decision of *whether* to evacuate is positively impacted by peer behavior, while the decision of *how* to evacuate is negatively impacted by peers. Furthermore, an increase in flood threat level lessens the magnitude of these impacts. These findings have important implications for the design of tailored emergency messaging strategies. Specifically, emphasizing or deemphasizing the severity of each threat in a multi-hazard scenario may assist in: (1) encouraging a reprioritization of competing risk perceptions and (2) magnifying or neutralizing the impacts of social influence, thereby (3) nudging evacuation decision-making toward a desired outcome.

**INTRODUCTION**

The novel coronavirus (COVID-19) pandemic has overlapped with numerous climate disasters in the United States, including floods, earthquakes, tornadoes, wildfires, and hurricanes (Smith, 2021). Prior research has shown that evacuation decision-making is primarily influenced by access to resources, risk perception, and social influence (Dash and Gladwin, 2007; Huang et al., 2012; Riad et al., 1999; Sadri et al., 2017; Sadri et al., 2021). However, this existing body of research has primarily focused on single-hazard events, while a research gap persists pertaining to the study of evacuation decision-making in the context of multi-hazard disasters. In addition to the many major flooding events that have occurred during the pandemic, including in Germany and China (Simonovic et al., 2021), this study is motivated by a flood evacuation of over 11,000 individuals from parts of Midland, Michigan and surrounding areas caused by the failure of two dams: the Edenville and downstream Sanford. This event occurred on May 19th - 20th, 2020 while the state was under a pandemic stay-at-home mandate. These dual climate and health hazards present contradictory strategies: sheltering-in-place to isolate oneself against viral exposure versus gathering with the masses on roads and in shelters to evade localized threats to life and property. Crowds are undesirable in both natural hazard emergencies and contagious health emergencies, and yet, social connectivity bolsters resilience to these same emergencies by providing access to material and immaterial resources. Therefore, it is difficult to predict what effect peer behavior will have on the decisions of whether and how to evacuate during a multi-hazard disaster.

The goal of this research is to better understand flood evacuation decision-making amid the viral COVID-19 pandemic. Specifically, this study examines the following three research questions:

1) What is the impact of *social influence* on multi-hazard evacuation decision-making? Using the observation of peers' evacuation decisions as a proxy for social influence, we investigate its mixed impact (i.e., following versus avoiding) on evacuation decision-making, depending on the type of evacuation decision (i.e., whether or how to evacuate) within the dual emergency context (i.e., a pandemic-concurrent flood).

2) What is the impact of *flood threat level* on multi-hazard evacuation decision-making? For example, a more severe emergency may lead to assigning a lower (or higher) importance to peer influence. Our work investigates the heterogeneity in these peer effects.

3) What is the impact of *pandemic risk perception* on multi-hazard evacuation decision-making? Here we home in on the way decision-makers navigate the competing role of flood threats and pandemic risks stemming from co-existing emergencies.

To investigate these questions, we collect survey data using a stated choice experiment to model hypothetical responses to a pandemic-concurrent flood evacuation. Discrete choice models in willingness-to-pay space with random parameters are developed to control for heterogeneity in evacuation decisions.

This paper is organized as follows. The next section presents a literature review, followed by an overview of the data collection and experimental design. Then we present the discrete choice methodology and discuss the results. The last section concludes with the implications of the findings.

**LITERATURE REVIEW**

The literature review summarizes previous studies on factors that influence evacuation decisions. Concluding each subsection, we outline how the previous research guided our hypotheses and modeling in the current work.

**Evacuation Decision-Making**

***Evacuation Decision and Mode Determinants***

Decision-making during evacuations is often closely related to sociodemographics. For example, the decision of *whether* to evacuate is typically found to be positively correlated with female gender, children at home, and education level, while negatively correlated with age, household size, and homeownership (Huang et al., 2016). Inability to evacuate is often related to race, income, disability, and health status (Renne et al., 2011). Additional behavioral factors influencing the decision of whether to evacuate include perceptions of risk, self-efficacy, and communication (such as official warnings, environmental cues from storm conditions, and social influence or connectivity), as well as income, race, employment status, duration of residence, and prior disaster experience (Collins et al., 2018; Demuth et al., 2016; Huang et al., 2016; Lazo et al., 2015; Metaxa-Kakavouli et al., 2018; Pei et al., 2020). It is worth noting that most of these studies focus on single-hazard evacuation scenarios.

The decision of *how* to evacuate is typically examined through studies on mode choice and route selection. Prior work suggests that a personal vehicle is the most commonly used evacuation mode (Lindell et al., 2011; Wong et al., 2018). Recently, research has begun to consider the role of ridesharing for evacuation of individuals who lack access to personal vehicles. *Single*-hazard evacuation research has found evidence of driver willingness to offer shared evacuation rides with strangers (Li et al., 2018; Wong and Shaheen, 2019). Recent *multi*-hazard evacuation research has shown that the COVID-19 pandemic is impacting evacuation decision-making and reducing evacuation likelihood (Alam and Chakraborty, 2021; Collins et al., 2021, Borowski et al., 2021).

Further research on multi-hazard evacuation mode choice is warranted. The current paper examines the decision of whether and how to evacuate, while considering social and emotional factors, as well as communication and perceptions of relative threat and risk levels.

### Social Influence, Networks, and Peer Effects

The influence of social networks on evacuation decision-making is an ongoing area of research. It has been shown that families, relatives, friends, neighbors, and coworkers impact evacuation decision-making (Gladwin et al., 2007). Contagion-based network science analysis has been used to simulate the impact of family relationships (e.g., parents, siblings, relatives, etc.), as well as other social ties (e.g., neighbors, colleagues, friends, etc.), on evacuation decision-making and cascading behaviors (Hasan and Ukkusuri, 2011). Typically, familial relationships have a stronger influence on evacuation decisions than community relationships (Perry, 1979). Several works account for the joint nature of evacuation decisions. This has been examined in relation to intervals of interactions, conversation contexts, and distance of social contacts (Sadri et al., 2015). Similarly, research has examined social influence on evacuation decision-making within the context of household gatherings (Liu et al., 2014). Additional research has provided support for the impact of social network structures on evacuation decision-making (Ahmed et al., 2020; Sadri et al., 2017).

For a comprehensive review of social influence on evacuation decisions, please refer to Sadri et al. (2021), which recommends future research on new models for encouraging shared mobility during extreme events to addresses the question of under what conditions evacuees are willing to share a ride. The present study begins to explore this research gap by including shared rides in our evacuation choice experiment.

### Emotionality

Emotional states have been shown to have a significant effect on decision-making (Chorus et al., 2013; Gutteling et al., 2018; Hess and Stathopoulos, 2013; Lerner and Keltner, 2000; Liu et al., 2017). Experiencing heightened emotions may lead to "hot-state" decision-making or impulsive actions in pursuit of one's visceral desires (Reid, 2010). In the context of evacuations, it is likely that negative emotions will predominantly impact decision-making. The four primary negative emotions that are typically studied in decision-making research are fear, anger, sadness, and anxiety (Jin, 2009; Jin et al., 2016; Kim and Cameron, 2011; Lerner and Keltner, 2000). It is important to note that although emotionality can impact decision-making, most of the research in this area has revealed that panicking is rarely observed during emergency events (Mileti and Peek, 2000; Quarantelli, 2001; Quarantelli and Dynes, 1977; Sorensen and Mileti, 1988).

The present study explores the impacts of fear and anxiety on evacuation decision-making in a multi-hazard scenario.

**Emergency Messaging: Risk and Threat Severity**

Research suggests that emergency messaging impacts decision-making through six stages of communication: hearing, confirming, understanding, believing, personalizing, and responding (e.g., Mileti and O'Brien, 1992; Mileti and Peek, 2000; Mileti and Sorensen, 1990). Beyond formal messaging, individuals often rely on social networks to gather information to support decision-making (Lindell et al., 2019). Research on the use of on-demand ridesourcing for evacuation has shown that this decision is impacted by the notification source and the level of urgency communicated (Borowski and Stathopoulos, 2020). The message content, style, and receiver characteristics (such as social setting, social ties, social structure, psychological factors, and pre-warning perceptions) can all impact decision-making and response (Mileti and Peek, 2000). The emotional content of messages can also affect decision-making, as observed for messages emphasizing impacts on buildings and property and those emphasizing impacts on human life, both of which were found to have a positive effect on evacuation intention, risk perception, and response efficacy (Morss et al., 2018).

The present study advances this line of research by examining the impact of evacuation messaging with increasing threat levels on the decisions of whether and how to evacuate in a multi-hazard scenario.

**Pandemic Impacts on Travel and Evacuation Behavior**

The COVID-19 pandemic has had a distinct impact on travel behavior, including less time spent commuting (U.S. Bureau of Labor Statistics, 2021) and a decline in shared mode ridership, such as public transit, subways, and bikesharing systems (Abdullah et al., 2020; De Vos, 2020; Rahimi et al., 2021; Teixeira and Lopes, 2020). Some research has begun to examine evacuation decision-making regarding whether to evacuate within the pandemic context. Findings show that, in general, pandemic concerns take precedence over flood concerns in a multi-hazard evacuation (Borowski et al., 2021; Botzen et al., 2021). Older individuals with greater vulnerability to COVID-19 are even more likely to stay behind during an evacuation (Botzen et al., 2021; Meng et al., 2020). This may be because the individuals at greatest risk of severe COVID-19 consequences are those 85 years of age and older, as well as those with underlying medical conditions, such as cancer, obesity, diabetes, serious heart conditions, and an immunocompromised state (Centers for Disease Control and Prevention, July 21, 2020). Regarding evacuation ridesharing during the pandemic, significant determinants include traditional factors, such as being Black, female, or a parent, as well as less traditional factors related to the sociopolitical context of the pandemic, including being Republican or a millennial (Borowski et al., 2021).

These significant factors influencing travel behavior within the context of the pandemic, such as age and ridesharing attitudes, are expected to be relevant to the multi-hazard evacuation event examined in this research.

## DATA COLLECTION

### Stated Preference Survey

A survey including a choice experiment was carefully designed to determine multi-hazard evacuation preferences with a focus on social influence, emotional response, and risk perception to study this multi-hazard decision-making phenomenon, as outlined in **Figure 1**. This survey was distributed online in three midwestern states (Illinois, Michigan, and Wisconsin) and three southern states (Georgia, Louisiana, and Mississippi) from June 30th through July 2nd, 2020. These states were chosen based on predictions of high flood risk (NOAA, 2020). Furthermore, each of the three states in each geographic region represented a different phase of pandemic restriction measures in July 2020 (i.e., reopened, reopening, and paused), as indicated in **Table 1**. The web-based survey was designed on Qualtrics (Qualtrics, 2005) and administered to 600 respondents using a Prolific panel (Prolific, 2014). After removing poor quality responses that failed an attention check question or showed patterns of inattentiveness, 586 respondents remained (98%). The final survey instrument included the following five sections:

1. **Emotion.** In section one, respondents received a description of the flood evacuation in Michigan that occurred in May 2020, including a quoted recollection of a real experience of the flood evacuation event. Respondents were asked how likely they would be to feel various emotions in that situation (i.e., angry, scared, anxious, and sad) as measured along a 5-point Likert-like scale.

2. **Pre-evacuation questions.** Respondents were asked about evacuation logistics, such as how many belongings they would take with them when evacuating, their household size, and where they would likely stay during a flood evacuation.

3. **Egocentric network name generator.** Here respondents were asked to list up to five individuals in their life who they expect would provide the most support in several resource domains (i.e., material, instrumental, emotional, and informational) during a flood evacuation. This section also included questions about these relationships regarding distance, duration, frequency, and similarity in terms of gender, ethnicity, and age.

4. **Evacuation discrete choice experiment.** Most relevant for the current paper is the choice experiment. The attribute selection and presentation were qualitatively evaluated and then underwent a pilot with respondents from Prolific (n = 30). Drawing on priors from the pilot, an efficient Bayesian design was created using Ngene software (ChoiceMetrics, 2021). A total of 27 hypothetical flood evacuation scenarios were varied across three survey blocks to reduce respondent fatigue. To anchor participants in the context, respondents were shown a map with a 5-mile radius that they would need to evacuate within 2 hours. They were then given a gradually changing evacuation notification indicating the flood threat level for each choice scenario. To control for the flood threat level, a color-coded warning message was displayed in three levels (low = green, moderate = yellow, extreme = red) for three scenarios each, as shown in **Figure 2**. The choice tasks contained three evacuation ride alternatives and the option to stay home for that scenario. The choice attributes for the three rides included the ride cost, walking distance to the vehicle, wait time prior to travel, travel time to the destination, and in-vehicle crowding (i.e., some rides were private, and others were shared with one stranger either in the front or back seat). To model social peer effects, for all four evacuation alternatives, respondents were told what percentage of their egocentric social network selected each alternative. The egocentric social network data was collected as described in the third point above. This method of measuring social influence is novel and inspired by Gaker et al. (2010). **Figure 3** shows a hypothetical scenario from this task.

5. **Sociodemographics.** In this section, respondents were asked about their personal sociodemographics, as well as household constraints (such as living with one or more children under the age of 5, one or more adults of age 65 years or older, one or more pets, etc.). The survey sample sociodemographics are shown in **Table 2**.

**METHODOLOGY**

**Random Parameter Model: Willingness-to-Pay Space**

In discrete choice modeling, the standard utility parameterization occurs in preference space (Train, 2009). In this paper, we also specify our model in willingness-to-pay space wherein the coefficients directly represent the respondents' willingness to pay (Train and Weeks, 2005). A benefit of this approach is its added flexibility in estimating random parameter distributions because willingness-to-pay parameters can take on any distribution chosen during the estimation.

The utility function $U$ for individual $n$ choosing alternative $j$ for choice task $t$ is given by:

$$U_{njt} = \beta'_n x_{njt} + \varepsilon_{njt} \tag{1}$$

We can respecify this model so that the coefficient estimates directly represent the willingness-to-pay estimation. This provides an alternative to the standard practice of dividing non-price attribute parameters by the price parameter to obtain the willingness-to-pay estimates. The advantage is that as willingness-to-pay estimates are directly defined for the parameter ratio, estimates are more tractable, plausible, and relevant for policymakers (dit Sourd, 2020; Mabit et al., 2006; Sonnier et al., 2007; Train and Weeks, 2005).

More formally, **Eq. 1** can be rewritten to represent respondents' willingness-to-pay space following Scarpa et al. (2008) and Train and Weeks (2005), the only difference being that the price $p$ is isolated from the non-price evacuation ride attributes. Here coefficient $\lambda_n$ is related to price and $\beta'$ represents all other attributes.

$$U_{njt} = -\lambda_n p_{njt} + \beta'_n x_{njt} + \varepsilon_{njt} \tag{2}$$

Given a scale parameter of $\mu$ where the error variance is expressed as $\mu_n^2 \left(\frac{\pi^2}{6}\right)$, **Eq. 2** can be divided by $\mu_n$.

$$U_{njt} = \left(\frac{-\lambda_n}{\mu_n}\right) p_{njt} + \left(\frac{\beta_n}{\mu_n}\right)' x_{njt} + \varepsilon_{njt} \tag{3}$$

Rewriting $\left(\frac{-\lambda_n}{\mu_n}\right) = \phi_n$ and $\left(\frac{\beta_n}{\mu_n}\right) = \xi_n$ and inserting into **Eq. 2** we obtain:

$$U_{njt} = \phi_n p_{njt} + \xi_n' x_{njt} + \varepsilon_{njt} \tag{4}$$

Calculating willingness to pay based on the preference space model in **Eq. 4** is executed as $z_n = \left(\frac{\xi_n}{\phi_n}\right)$ and thereby $\xi_n = z_n \phi_n$. In practice, if we specified $\phi$ and $\xi$ as randomly distributed parameters that vary over decisionmakers following a given distribution, the willingness-to-pay calculation needs to contend with two random terms (Daly et al., 2012).

Re-parameterizing **Eq. 4** gives us the willingness-to-pay space model in **Eq. 5**. This model is behaviorally equivalent to the preference space specification in **Eq. 4** (Train and Sonnier, 2005), but the distribution of $\phi_n$ affects the estimation of willingness-to-pay values compared to **Eq. 4**. In particular, if we assume normal distributions, we may contend with highly skewed or unidentified moments of willingness to pay (Daly, 2012).

$$\begin{cases} U_{njt} = \phi_n p_{njt} + (z_n \phi_n)' x_{njt} + \varepsilon_{njt} \\ U_{njt} = \phi_n [p_{njt} + z_n' x_{njt}] + \varepsilon_{njt} \end{cases} \tag{5}$$

Indeed, many of the standard assumptions of random parameter distributions can lead to unidentified moments in the preference space form.

We include random parameters in our model to account for taste heterogeneity across respondents by allowing these parameters to vary by individual. The random parameter coefficients can be expressed as the sum of the population mean and the stochastic deviation representing each respondents' tastes, experiences, and uncertainty with that parameter while controlling for other effects in the model (Train, 2009). For our evacuation modeling, we include random terms for price and peer effects, and after comprehensive comparison, we found that the willingness-to-pay space formulation provides better fit and more plausible results.

In our preference space model, *Cost* is specified as a random parameter with a log-normal distribution, which ensures a strictly negative sign, in line with many prior works (e.g., Hole-Kolstad, 2012; Kjaer et al., 2013; Matthews et al., 2017). After systematic testing, two additional random parameters are found to be significant for the social effects parameters: the *Share of Peers Staying* and the *Share of Peers Choosing a Ride*. These parameters are both defined as normally distributed to reflect the range of sensitivities associated with peer effects, spanning from avoiding to following behaviors. Interactions between *Peer Choice*, *Flood Threat*, and *Pandemic Risk Perception* are also estimated.

For our data analysis, we first estimated preference space models to identify significant fixed and random parameters, as well as interactions. After the preference space model estimation was complete, we specified the same model in willingness-to-pay space. We used the Apollo package for R (Hess and Palma, 2019) to estimate the model in both spaces, employing the Broyden–Fletcher–Goldfarb–Shanno (BFGS) algorithm. For the random parameter estimation simulation, 1000 Halton draws were found to produce stable parameter estimates.

**Table 3** lists the significant variables that were included in the preference space and willingness-to-pay space model estimations. In our discussion, we will refer to the coefficients estimated in willingness-to-pay space, but we also include the preference space estimation for comparison and to highlight the improvement of fit obtained using willingness-to-pay estimation.

It is worth noting that willingness-to-pay coefficients directly provide a monetary value for each modeled parameter. For the decision of *how* to evacuate, our study examines shared rides and standard mode attributes, such as cost and travel time. In particular, the monetary value of ride attributes could have a direct interpretation for the design of emergency ridehailing services, including recommended discounts, promotions, and reimbursements.

**RESULTS**

The results of the random parameter logit model estimated in preference space and willingness-to-pay space are shown in **Table 4**. The willingness-to-pay space estimation improves the fit of the preference space estimation with an increase in final log-likelihood of 220 and a higher adjusted rho squared. Three random parameters, twelve fixed parameters, and five interactions are included in the final model. All are significant to a 95% level of confidence or better except for the parameter *Pets*, which is insignificant in the willingness-to-pay estimation. The following subsections focus on the willing-to-pay estimation results related to the random parameters for ride cost and social influence, as well as the fixed parameters for ride attributes, threat level, risk perception, and emotion.

**Random Parameters**

In the process of building our model, we opted to keep the random coefficients that had statistically significant standard deviations. It is informative to compare the distribution of the random parameters in the two model structures. In the preference space estimation shown in **Figure 4 (a – c)**, the cost parameter distribution is narrower with a coefficient of variation (COV) for *Cost* that is relatively small (0.28), while that for *Share of Peers Choosing a Ride* and *Share of Peers Staying* are both relatively large (1.50 and 1.40, respectively). This suggests low variance (or general agreement across respondents) regarding the impact of *Cost* on selecting a ride and high variance (or general disagreement across respondents) regarding the impact of the share of peers on the decisions of whether and how to evacuate.

The distribution of the random parameters in the willingness-to-pay space shown in **Figure 4 (d – f)** reveal less dispersion in each case (COV for *Cost* is 0.17, *Share of Peers Choosing a Ride* is 0.80, and *Share of Peers Staying* is 0.21). This suggests low variance in that the direct ratio estimation is better at capturing the heterogeneity in preferences for these factors, while we note that there is higher variance for the impact of *Share of Peers Choosing a Ride*.

In other words, in both models, respondents displayed the greatest diversity in their views of the benefits and risks associated with "following the crowd" in an evacuation setting.

*Social Influence and Flooding Threat*

Willingness to pay levels out when it relates to social influence and flood threat as shown in **Figure 5**. The strongest effects are observed for the *Share of Peers Staying* estimates. To aid interpretation, the results suggest that if the proportion of peers staying increases from 0% to 100%, respondents would be willing to pay $271.13 on average to follow their peers' decision to stay, which resembles earlier research findings on decision inertia across consecutive hurricane evacuation events (Murray-Tuite et al., 2012). This finding suggests a dominant effect of peer behavior when it comes to staying despite a flood warning. This coefficient is 3.7 times the magnitude of that for *Share of Peers Choosing a Ride*, suggesting a greater resistance to changing behavior related to staying compared to the mode choice effect. The context matters, however. When the *Flood Threat* is upgraded to moderate, this willingness-to-pay amount decreases to $43.89 (i.e., $271.13 – $227.24), and when *Flood Threat* is extreme, this amount switches sign, expressing the need to be compensated with $76.84 (i.e., $271.13 – $347.97) to stay. Overall, we note that the impacts of the *Flood Threat* interactions with *Share of Peers Staying* are about 5 times greater than those with *Share of Peers Choosing a Ride*, suggesting that the impact of flood threat communication on the decision of *whether* to evacuate is greater than on the decision of *how* to evacuate.

Looking at the evacuation mode choice, the finding can be interpreted as follows: if the *Share of Peers Choosing a Ride* increases from 0% to 100%, respondents would need to be compensated an additional $71.77 on average to follow their peers. When the *Flood Threat* is elevated to moderate, this compensation amount decreases to $28.17 (i.e., $71.77 – $43.60), and when *Flood Threat* is extreme, this amount instead suggests a willingness to pay $1.04 (i.e., $71.77 – $72.81).

**Fixed Parameters**

*Ride Attributes*

The flooding experiment included five mobility service attributes, in addition to the peer effect and threat level attributes discussed above and shown in **Figure 2.** The willingness to pay for each of the ride attributes is shown in **Figure 6**. Starting with the time attributes, for every additional minute of *Travel Time*, respondents need to be compensated $0.63 to accept a ride, while for every additional minute required to *Wait* for the evacuation ride to arrive, respondents need to be compensated $0.79. These findings suggest that the disutility of waiting time is 125% that of traveling time, meaning that waiting is perceived to be more costly than traveling time in an emergency. This reflects a similar asymmetry observed for non-emergency applications (Arentze and Molin, 2013). For every additional mile required to *Walk* to access a ride during a flood evacuation, respondents need to be compensated $32.78 to accept that ride, suggesting that walking is perceived as onerous in the emergency setting. This reflects prior research findings that the feasibility of walking during a flood evacuation is a function of water depth (Dias et al., 2021; Liu et al., 2009). To evacuate in the *Backseat* of a shared ride, respondents need to be compensated $36.74, while evacuating in the *Front Seat* of a shared ride, respondents need to be compensated $27.72.

*Sociodemographics*

Willingness to pay as it relates to stay determinants is shown in **Figure 7**. For every pound increase in *Evacuation Luggage*, respondents need to be compensated $4.36 to evacuate. For every additional ten years of *Age*, respondents need to be compensated $5.75. Respondents in households with one or more *Pets* need to be compensated $7.54. Respondents with a *Disability* need to be compensated $84.97. The magnitude of this value of willingness-to-pay is approximately 11 times greater than that for the other binary parameter, *Pets*, suggesting a greater severity of this evacuation constraint. In the course of specification testing, other factors, including gender, age, race/ethnicity, median household income, home ownership, vehicle ownership, residential area type, and political affiliation, were tested but resulted as being insignificant in the final model.

*Risk Perception and Emotion*

Risk perception is one of the key elements to understand how households make decisions on when to evacuate. Earlier work shows that risk is not perceived in the same way for all decisionmakers and may be influenced by social dimensions and the evacuation context (Dash and Gladwin, 2007). Perceived risk for flooding has been shown to be particularly important in a flooding context (Whitehead et al., 2000). Emotion has been shown to influence evacuation decision-making by impacting risk perception and message interpretation (DeYoung et al., 2019; Slovic and Peters, 2006). Our model results illustrate the impact of subjective risk perceptions and evacuation severity against the emotions experienced by respondents.

Out of the four hypothesized emotional responses to the flooding evacuation storyline, only two were shown to be impactful, namely anxiety and fear. In terms of interpretation, for each additional level of *Evacuation Anxiety* likelihood (5-point scale), respondents need to be compensated $11.17 to evacuate. Instead, for each additional level of *Evacuation Fear* likelihood (5-point scale), respondents are willing to pay $14.74 to evacuate.

The survey also controlled for the role of COVID-19 pandemic contagion concerns, in the form of a scale question. Results show that respondents who view the pandemic as a *Major Risk* to their health (binary variable) need to be compensated $219.45 to evacuate. There was no evidence of effects below the level of major risk (i.e., moderate, minor, or none), suggesting that a high threshold exists for pandemic risks in this evacuation setting. Once reached, however, the impact seems to surpass emotional effects. There is an important interaction effect between pandemic risk and flood threat communication. For each additional level of *Flood Threat* (4-point scale), the pandemic concern reduces by the equivalent amount of $85.02.

**DISCUSSION AND PRACTICAL IMPLICATIONS**

**Nuanced Impact of "Following the Crowd"**

Prior research on single-hazard evacuation scenarios suggests that social influence tends to result in a follow-the-crowd mentality (Sadri et al., 2021). This may be due to the sense of safety-in-numbers that the act of following others could provide (Lindell et al., 2005). However, in a multi-hazard scenario that involves both a climate and public health crisis, it is unclear whether a similar effect of social influence would be observed. In this type of multi-hazard event, the emergency messaging tends to be contradictory. Specifically, the advice provided for one crisis suggests staying home (the pandemic), while the advice for the other crisis suggests evacuating (flooding).

Our findings show that although the impact of social influence differs widely across respondents, overall, it is *negative* for ride selection and *positive* for the decision to stay. In other words, if more peers choose to stay despite official flood warning, the respondent will be more likely to stay, as well. Instead, if more peers choose a given evacuation ride, respondents will be more likely to choose a different ride option. This makes sense given the pandemic-related advice to shelter-in-place and to follow social distancing protocol when leaving one's home.

When interacting social influence with flood threat messaging, two important observations can be made. First, when the threat communication is upgraded, evacuation scenarios with greater flood threat result in reduced impacts of social influence. That is, for the decision to stay, the original mean value of the willingness to pay to follow peers (i.e., $271.13) is reduced by 84% with moderate flood risk and switches to a need to be compensated to stay in scenarios with extreme flood risk. This indicates that flood risk communication seems to supersede the reliance on following crowd cues, at least for severe events. Second, more broadly, this interaction effect also suggests that as the severity of the flood threat increases, respondents prioritize avoiding the flood risk over avoiding the pandemic risk.

For ride selection, the original need to be compensated $71.77 to follow peers' ride choice is reduced by 61% to $28.17 in scenarios with moderate flood threat and switches to a minimal willingness to pay of just $1.04 to follow peers' ride selection in scenarios with extreme flood threat. This increased desire to follow peers' ride selection in evacuation scenarios with greater threat of flood risk may be related to perceiving the consequences of selecting an ineffective evacuation ride as being more severe and therefore wanting to minimize any regret of making a poor choice in an evacuation scenario with extreme flood threat.

The magnitude of the impact of the share of peers staying is 3.8 times greater than that of the share of peers choosing a ride, although the magnitude of the standard deviation for each is approximately equivalent. This asymmetry of social influence may be related to the signaling effect of visible peer behavior. The evacuation behavior of others could provide an environmental cue emphasizing the urgency of the scenario. When individuals evacuate, this behavior is more visible than their decision to stay and could result in a springing into action that motivates others to evacuate, as well.

**The Dueling Risk Perceptions and Emotions**

Our findings show that different emotional reactions lead to contrasting effects. For every level of increased evacuation anxiety, respondents are willing to pay $11.17 to stay during an evacuation. By contrast, for every level of increased evacuation fear, respondents would require compensation of $14.74 to stay during an evacuation. One possible explanation for these divergent findings is that anxiety may lead to a "freeze" response, while fear may lead to a "flee" response. The impact of fear is greater in magnitude than the impact of anxiety by 132%.

This research also offers insight on the tension between pandemic concerns that suggest avoiding exposure introduced by evacuating in shared vehicles and flooding destruction that impels respondents to leave. Respondents who view the pandemic as a major risk to their health are willing to pay a high price to stay during an evacuation, but for every increase of flood threat level, that amount reduces notably. Given the three levels of flood threat in the experiment (i.e., low, moderate, and extreme), this reduction could be so great as to entirely counterbalance the effect of pandemic risk evaluation. In practical terms, this suggests that respondents who view the pandemic as a major threat could demonstrate a need to be compensated to stay if the flood threat was great enough.

**Emergency Communication**

The findings of this study have important implications for multi-hazard emergency communication. Pandemic-concurrent evacuation decision-making can be nudged in different directions by emphasizing different hazard components and their associated threat levels, which can impact the magnitude of the social influence effect. Peer behavior is largely outside of the control of emergency communication management, but some behaviors could be broadcasted and amplified, such as communicating the fact that others are evacuating. Furthermore, results from our analysis indicate there is an inherent asymmetric signaling effect that occurs, likely due to the decision to evacuate being more visible to others compared to the decision to stay. This observation presents a potential two-pronged approach to communicating peer behavior via both formal and informal channels. Communication strategies should utilize risk perception, emotion, and behavioral broadcasting carefully to intentionally nudge evacuation decision-making toward the desired outcome.

**SUMMARY AND CONCLUSION**

In this paper, we analyze the decisions of whether and how to evacuate during a flooding disaster occurring simultaneously with the global COVID-19 health emergency. The focus of the analysis is on social influence and how it is affected by emergency messaging, emotionality, and the overlapping hazard setting. This problem is challenging given that the multi-hazard setting will likely necessitate a nuanced navigation of competing decision-making strategies wherein a desire to follow peers is weighed against contagion risks.

Drawing on data from a stated choice experiment, we report the results of a random parameter logit model in willingness-to-pay space. We examine the decision of whether to evacuate or stay at home in response to a flooding event. The modeling approach offers flexibility in assessing both unobserved and systematic heterogeneity present in evacuation responses. The critical takeaway from this research is that social influence during a multi-hazard evacuation event has vastly different impacts depending on the decision being made, threat levels, and taste heterogeneity. Overall, our findings show that social influence has a positive impact on *whether* one evacuates (i.e., stay versus go) and a negative impact on *how* one evacuates (i.e., evacuation mode choice). As flood threat increases, the magnitude of the effect of social influence decreases. Large taste heterogeneity across respondents regarding the impact of social influence is observed. Finally, emotion is found to be significantly correlated with evacuation decision-making, but only for fear and anxiety. Specifically, evacuation anxiety is positively correlated with the decision to stay, while evacuation fear is positively correlated with the decision to evacuate.

There are three main limitations to this work. First, there is a demographic imbalance in the survey sample that is skewed toward individuals under the age of 45, those who are Asian, and those of higher income. This is due in part to using convenience-based sampling, which is justified for evacuation research (e.g., Wong et al., 2020) and COVID-19 travel behavior studies (e.g., Parady et al., 2020) due to safety concerns. Second, the results were found using a stated preference choice experiment, which presents the risk of hypothetical bias. Lastly, in this exploratory work, the emotion variables (i.e., anxiety and fear) were each measured using a single survey question and modeled as continuous variables, while more accurately, emotion should be treated as a latent factor using multiple indicators. Future work should extend the analysis of the impact of emotion on multi-hazard evacuation decision-making by treating it as a latent variable.

## DATA AVAILABILITY STATEMENT

Some data, models, or code generated or used during the study are proprietary or confidential in nature and may only be provided with restrictions. Survey data: with IRB approval permission from Northwestern University. Some data, models, or code that support the findings of this study are available from the corresponding author upon reasonable request. Models: available within this paper. Model code: available from the corresponding author upon reasonable request.

## ACKNOWLEDGEMENTS

This research was supported in part by funding from the National Defense Science and Engineering Graduate (NDSEG) fellowship provided to Elisa Borowski and the U.S. National Science Foundation (NSF) Career grant No. 1847537 and from Leslie and Mac McQuown via Northwestern University Center for Engineering Sustainability and Resilience Seed Funding to Amanda Stathopoulos. The survey is approved by Northwestern's IRB with study number STU00211228.

**TABLES**

**Table 1.** COVID-19 statistics for July 2020 in surveyed U.S. states

| State | New Cases | 7-Day Avg | New Deaths | 7-Day Avg | Total cases | Total deaths | Order Began | Order Ended | Status During Survey | Governor |
|-------|-----------|-----------|------------|-----------|-------------|--------------|-------------|-------------|---------------------|----------|
| Georgia | 2,309 | 1,900 | 21 | 17 | 124,267 | 3,071 | 4/3/20 | 4/30/20 | Reopening | Republican |
| Illinois | 880 | 788 | 24 | 25 | 160,898 | 7,468 | 3/21/20 | 5/29/20 | Reopening | Democrat |
| Louisiana | 2,083 | 1,099 | 17 | 12 | 88,700 | 3,509 | 3/23/20 | 5/15/20 | Pausing | Democrat |
| Michigan | 429 | 367 | 5 | 12 | 80,759 | 6,358 | 3/24/20 | 6/1/20 | Pausing | Democrat |
| Mississippi | 652 | 639 | 9 | 10 | 40,829 | 1,332 | 4/3/20 | 4/27/20 | Reopened | Republican |
| Wisconsin | 584 | 522 | 1 | 4 | 44,181 | 841 | 3/25/20 | 5/13/20 | Reopened | Democrat |

**Table 2.** Survey sample sociodemographics

|  | Survey | Georgia | Illinois | Louisiana | Michigan | Mississippi | Wisconsin |
|--|--------|---------|----------|-----------|----------|-------------|-----------|
| **Residence** | | | | | | | |
| United States | | 3.2% | 4.0% | 1.4% | 3.1% | 0.9% | 1.8% |
| Georgia | 23.4% | | | | | | |
| Illinois | 30.5% | | | | | | |
| Louisiana | 7.1% | | | | | | |
| Michigan | 15.9% | | | | | | |
| Mississippi | 4.1% | | | | | | |
| Wisconsin | 13.4% | | | | | | |
| **Gender** | | | | | | | |
| Male | 46.3% | 48.6% | 49.1% | 48.8% | 49.0% | 48.5% | 49.4% |
| Female | 51.9% | 51.4% | 50.9% | 51.2% | 51.0% | 51.5% | 50.6% |
| Other | 1.0% | | | | | | |
| **Age** | | | | | | | |
| 18 - 24 | 30.1% | 9.9% | 9.3% | 9.4% | 9.6% | 10.2% | 9.4% |
| 25 - 34 | 31.4% | 13.7% | 13.9% | 13.8% | 12.9% | 12.7% | 12.7% |
| 35 - 44 | 16.2% | 13.3% | 12.9% | 12.6% | 11.6% | 12.8% | 12.1% |
| 45 - 54 | 11.9% | 13.3% | 12.8% | 12.2% | 12.9% | 12.1% | 12.7% |
| 55 - 65 | 6.8% | 12.2% | 13.1% | 12.9% | 14.0% | 12.8% | 14.2% |
| 65 + | 3.5% | 13.8% | 15.6% | 15.5% | 17.2% | 15.9% | 17.0% |
| **Race** | | | | | | | |
| White | 65.0% | 58.2% | 71.7% | 61.8% | 78.3% | 58.1% | 85.3% |
| African American | 14.5% | 31.2% | 13.8% | 32.2% | 13.6% | 37.8% | 6.3% |
| Asian | 13.6% | 4.1% | 5.6% | 1.6% | 3.2% | 0.9% | 2.8% |
| American Indian | 1.0% | 0.2% | 0.1% | 0.5% | 0.5% | 0.4% | 0.8% |
| Two or more | 3.3% | 2.2% | 2.0% | 2.0% | 2.5% | 1.4% | 0.5% |
| Other | 0.8% | 2.5% | 5.4% | 1.2% | 1.0% | 0.9% | 2.0% |
| **Income** | | | | | | | |
| < $10k | 8.6% | 11.1% | 11.2% | 11.9% | 13.0% | 12.6% | 12.2% |
| $10k to $20k | 8.8% | 13.2% | 11.5% | 15.6% | 13.4% | 16.1% | 11.3% |
| $20k to $30k | 10.4% | 15.6% | 13.6% | 16.1% | 14.0% | 16.6% | 13.1% |
| $30k to $40k | 7.8% | 13.5% | 12.3% | 11.6% | 13.4% | 15.3% | 14.0% |
| $40k to $50k | 9.4% | 10.4% | 10.2% | 10.6% | 10.1% | 12.0% | 13.0% |
| $50k to $60k | 9.8% | 8.2% | 8.8% | 8.6% | 8.4% | 8.0% | 10.2% |
| $60k to $80k | 12.1% | 11.1% | 12.3% | 11.1% | 11.5% | 9.5% | 12.6% |
| $80k to $100k | 8.8% | 5.9% | 7.1% | 5.4% | 6.2% | 3.8% | 5.9% |
| $100k to $120k | 6.4% | 3.8% | 4.6% | 3.4% | 3.7% | 2.4% | 3.1% |
| $120k to $150k | 6.3% | 2.7% | 3.3% | 2.5% | 2.7% | 1.4% | 2.0% |
| $150k to $200k | 2.8% | 2.1% | 2.3% | 1.5% | 1.7% | 1.0% | 1.2% |
| > $200k | 4.1% | 2.4% | 2.9% | 1.6% | 1.8% | 1.3% | 1.6% |

**Table 3.** Definitions of model variables

| | Variable | Definition | Minimum | Maximum | Mean | Standard Deviation |
|---|---|---|---|---|---|---|
| **Random Parameters** | | | | | | |
| Cost | Count | Monetary cost of selecting evacuation ride | 0 | 40 | 11.60 | 15.65 |
| Peer fraction choosing ride | Count | Fraction of social network selecting a ride alternative (out of 5) | 0 | 1 | 0.27 | 0.29 |
| Peer fraction staying | Count | Fraction of social network selecting the stay alternative (out of 5) | 0 | 0.8 | 0.19 | 0.22 |
| **Ride Attributes** | | | | | | |
| Travel time | Count | Time of evacuation ride to travel from origin to destination | 0 | 60 | 25.96 | 23.83 |
| Wait time | Count | Time to wait until evacuation ride arrives | 0 | 60 | 18.49 | 22.29 |
| Walking distance | Count | Distance to walk to access evacuation ride | 0 | 0.5 | 0.18 | 0.21 |
| Shared back seat | Binary | Indicator of riding in the backseat in a shared evacuation ride with a stranger | 0 | 1 | 0.05 | 0.22 |
| Shared front seat | Binary | Indicator of riding in the front seat in a shared evacuation ride with a stranger | 0 | 1 | 0.07 | 0.25 |
| **Threat Level** | | | | | | |
| Extreme evacuation threat | Binary | Indicator of evacuation notification stating an extreme flood threat to life and property | 0 | 1 | 0.33 | 0.47 |
| Moderate evacuation threat | Binary | Indicator of evacuation notification stating a moderate flood threat to life and property | 0 | 1 | 0.33 | 0.47 |
| **Sociodemographics** | | | | | | |
| Age | Categorical | Decade categories of respondent's age | 1 | 6 | 2.44 | 1.37 |
| Belongings | Count | One-tenth of pounds of luggage brought during evacuation | 0 | 7.5 | 3.45 | 1.99 |
| Disability | Binary | Indicator of respondent having a disability | 0 | 1 | 0.03 | 0.16 |
| Pets | Binary | Indicator of respondent living in a household with one or more pets | 0 | 1 | 0.49 | 0.50 |
| **Attitudinal Variables** | | | | | | |
| Evacuation anxiety | Likert scale | 5-point level of evacuation anxiety | 1 | 5 | 4.53 | 0.93 |
| Evacuation fear | Likert scale | 5-point level of evacuation fear | 1 | 5 | 4.37 | 1.00 |
| Major pandemic risk | Binary | Indicator of respondent viewing the pandemic as a major risk to their health | 0 | 1 | 0.39 | 0.49 |

**Table 4.** Random parameter logit model results

| | Preference Space | | | Willingness-to-Pay Space | | |
|---|---|---|---|---|---|---|
| | Estimate | Robust t-ratio | p-value | Estimate | Robust t-ratio | p-value |
| "Evacuate" Alternative Specific Constant | 4.225 | 8.638 | 0.000 | 4.200 | 8.613 | 0.000 |
| "Stay" Alternative Specific Constant | (fixed) | | | (fixed) | | |
| **Random Parameters** | | | | | | |
| Cost (mean) | -3.714 | -36.437 | 0.000 | -3.536 | -46.499 | 0.000 |
| Cost (stdv) | 1.028 | 12.610 | 0.000 | 0.605 | 8.328 | 0.000 |
| Share of Peers Choosing Ride (mean) | -1.273 | -5.095 | 0.000 | 71.771 | 6.449 | 0.000 |
| Share of Peers Choosing Ride (stdv) | -1.911 | -12.703 | 0.000 | 57.285 | 8.800 | 0.000 |
| Share of Peers Staying (mean) | 7.756 | 9.327 | 0.000 | -271.125 | -8.034 | 0.000 |
| Share of Peers Staying (stdv) | 10.872 | 9.566 | 0.000 | -57.813 | -2.717 | 0.003 |
| **Ride Attributes** | | | | | | |
| Travel Time | -0.020 | -11.613 | 0.000 | 0.634 | 8.309 | 0.000 |
| Wait Time | -0.026 | -14.651 | 0.000 | 0.792 | 12.409 | 0.000 |
| Walking Distance | -1.079 | -7.414 | 0.000 | 32.783 | 6.591 | 0.000 |
| Shared Ride (Back Seat) | -0.797 | -10.033 | 0.000 | 36.738 | 8.858 | 0.000 |
| Shared Ride (Front Seat) | -0.840 | -10.879 | 0.000 | 27.716 | 8.693 | 0.000 |
| **Sociodemographics** | | | | | | |
| Age | 0.213 | 3.874 | 0.000 | -5.745 | -2.337 | 0.010 |
| Evacuation Luggage | 0.101 | 2.458 | 0.007 | -4.363 | -2.615 | 0.004 |
| Disability | 1.132 | 2.551 | 0.005 | -84.968 | -2.492 | 0.006 |
| Pets | 0.426 | 2.678 | 0.004 | -7.536 | -1.361 | 0.087 |
| **Attitudinal Variables** | | | | | | |
| Evacuation Anxiety | 0.372 | 2.966 | 0.002 | -11.174 | -2.087 | 0.018 |
| Evacuation Fear | -0.330 | -2.874 | 0.002 | 14.735 | 3.368 | 0.000 |
| Major Pandemic Risk Perception | 3.328 | 9.266 | 0.000 | -219.452 | -7.076 | 0.000 |
| **Interactions** | | | | | | |
| Peers Choosing Ride × Moderate Flood Threat | 0.499 | 1.788 | 0.037 | -43.598 | -4.088 | 0.000 |
| Peers Choosing Ride × Extreme Flood Threat | 1.108 | 4.147 | 0.000 | -72.809 | -6.511 | 0.000 |
| Peers Staying × Moderate Flood Threat | -9.267 | -7.675 | 0.000 | 227.238 | 7.220 | 0.000 |
| Peers Staying × Extreme Flood Threat | -21.127 | -8.920 | 0.000 | 347.974 | 9.071 | 0.000 |
| Flood Threat × Major Pandemic Risk Perception | -1.483 | -10.094 | 0.000 | 85.021 | 7.970 | 0.000 |
| **Model Parameters** | | | | | | |
| Number of Draws | 1000 | | | 1000 | | |
| Type of Draws | Halton | | | Halton | | |
| Number of Individuals | 586 | | | 586 | | |
| Number of Modeled Outcomes | 5274 | | | 5274 | | |
| Final Log-Likelihood | -5645.22 | | | -5425.07 | | |
| Adjusted Rho-Square | 0.225 | | | 0.255 | | |
| **Coefficient of Variation** | | | | | | |
| Cost | -0.277 | | | -0.171 | | |
| Share of Peers Choosing Ride | 1.501 | | | 0.798 | | |
| Share of Peers Staying | 1.402 | | | 0.213 | | |

**FIGURES (uploaded as separate files)**

**Fig. 1.** Pandemic-concurrent flood evacuation survey content flow chart.

| COVID-19 SURVEY | | | |
|---|---|---|---|

| PANDEMIC EXPERIENCE | | | |
|---|---|---|---|
| IMPACTS | | ATTITUDES | |
| Health<br>Relationships | Social Isolation<br>Natural Disasters | Vaccination<br>Risk | Preventative Behaviors<br>Policies |

| SOCIAL NETWORK | | | |
|---|---|---|---|
| NAME GENERATOR | | PLACE GENERATOR | |
| Size<br>Role<br>Duration | Proximity<br>Communication Mode<br>Homophily (age, race, gender) | Places (Quantity)<br>People (Quantity) | Conversation (Quantity)<br>Conversation (Length) |

| MOBILITY BEHAVIOR | | | |
|---|---|---|---|
| PHYSICAL | | VIRTUAL | |
| Activities (Frequency)<br>Shared Modes<br>Third Places | Needs<br>Outdoor Public Spaces | Activities (Satisfaction)<br>Technologies<br>Needs | Challenges<br>Downsides<br>Gains, Losses, Barriers |

| PERSONALITY | | | |
|---|---|---|---|
| Extroversion vs. Introversion | Self-Efficacy | | Social Isolation |

| SOCIODEMOGRAPHICS | | | | |
|---|---|---|---|---|
| Residential Area<br>Vehicles<br>Home Office<br>Internet Access | Education<br>Employment<br>Essential Worker<br>Marital Status | Caretaker<br>Household Size<br>Children | Income<br>Race<br>Ethnicity | Gender<br>Sexual Orientation<br>Political Ideology |

**Fig. 2.** Threat level context associated with evacuation scenarios.

| A LOW THREAT TO LIFE AND PROPERTY FROM FLASH FLOODS | A MODERATE THREAT TO LIFE AND PROPERTY FROM FLASH FLOODS | AN EXTREME THREAT TO LIFE AND PROPERTY FROM FLASH FLOODS |
|---|---|---|
| There is a **low likelihood** (6% to 15% probability) of flooding rain with storms capable of minor flooding | There is a **moderate likelihood** (16% to 25% probability) of flooding rain with storms capable of minor flooding | There is a **high likelihood or greater** (26% probability or greater) of flooding rain with storms capable of moderate flooding |

**Fig. 3.** Hypothetical evacuation choice scenario example.

| | **RIDE 1** | **RIDE 2** | **RIDE 3** | **STAY** |
|---|---|---|---|---|
| **COST** | $20 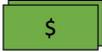 | $40 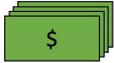 | $20 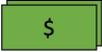 | |
| **WALKING DISTANCE** | 0 miles 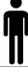 | 0.5 miles 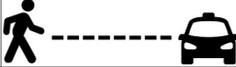 | 0 miles 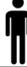 | |
| **WAIT TIME** | 60 mins 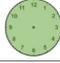 | 60 mins 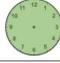 | 30 mins 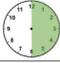 | |
| **TRAVEL TIME** | 20 mins 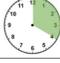 | 40 mins 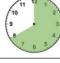 | 60 mins 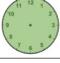 | |
| **CROWDING** | 4 ft 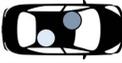 | Private 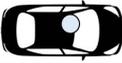 | Private 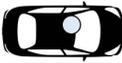 | |
| **OTHERS' ACTIONS** | 1 peer 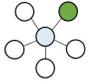 | 2 peers 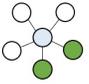 | 0 peers 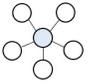 | 2 peers 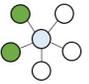 |

**Fig. 4.** Random parameter distributions for cost and social peer effects.

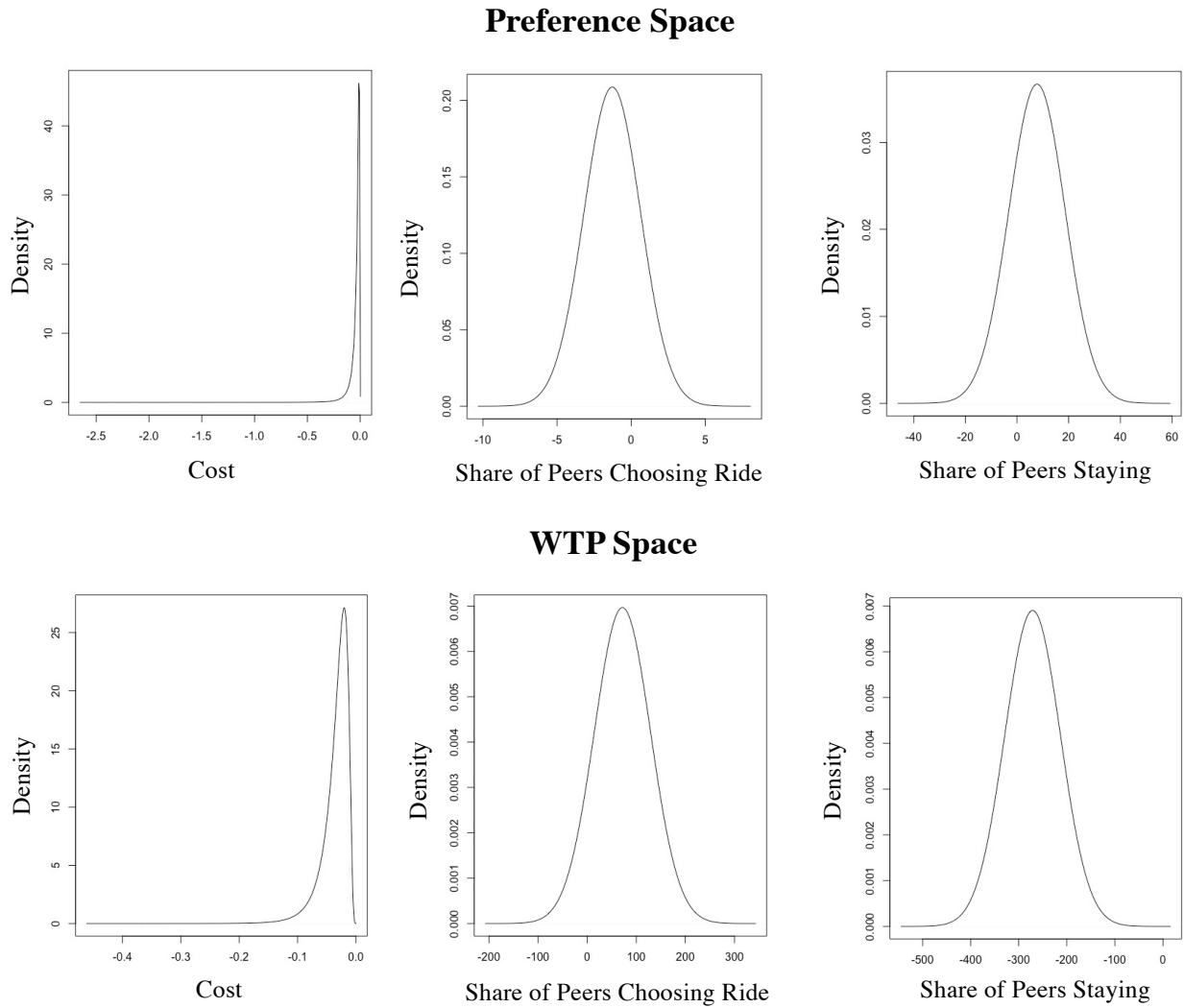

Notes for figure 4:

**Upper panel shows** *preference space* **(a)** Distribution of *Cost* random parameter, and distribution of willingness to follow peers' **(b)** evacuation *Ride* selection or **(c)** decision to *Stay* during an evacuation across all draws.
**Lower panel shows** *willingness-to-pay space* **(d)** Distribution of *Cost* random parameter, and distribution of willingness-to-pay to follow peers' **(e)** evacuation *Ride* selection or **(f)** decision to *Stay* during an evacuation across all draws.

**Fig. 5.** Willingness to pay to follow peers.

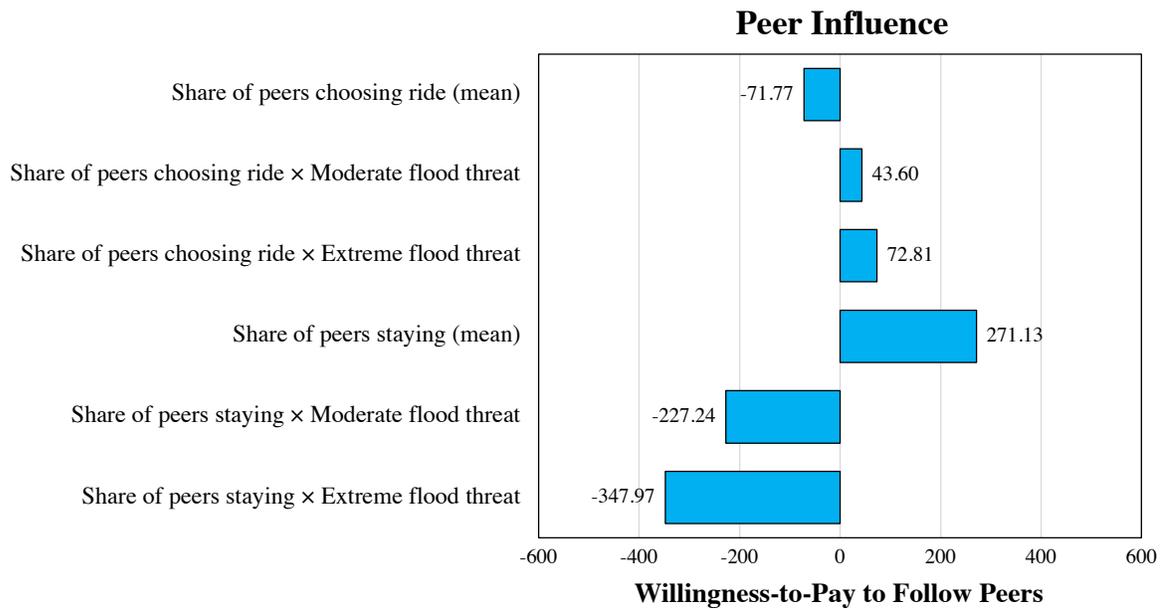

**Fig. 6.** Willingness to pay to choose an evacuation ride.

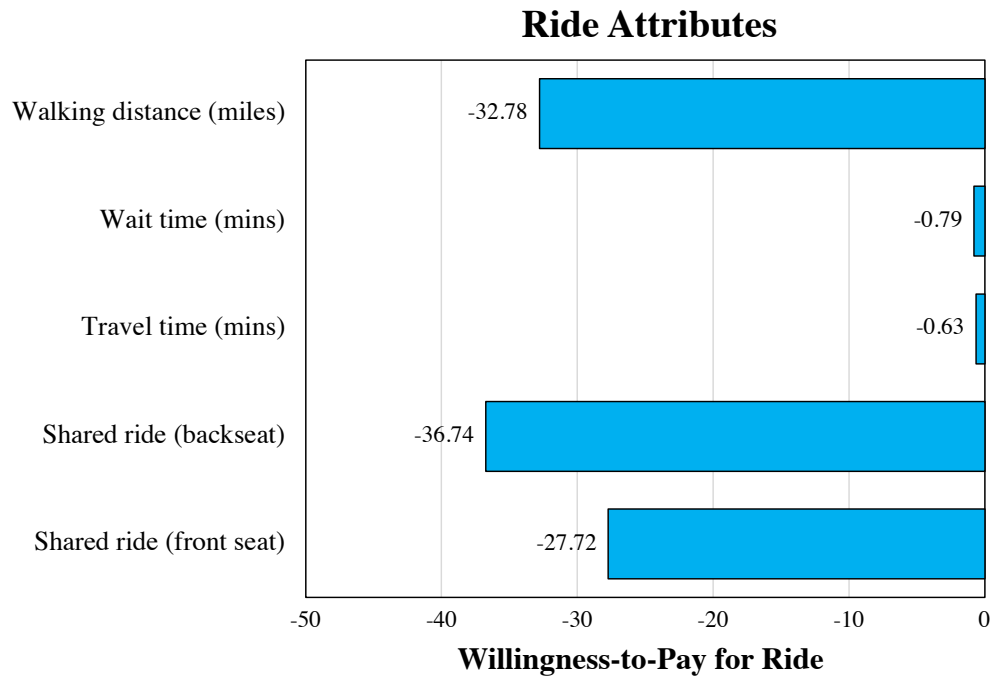

**Fig. 7.** Willlingness to pay to stay factors.

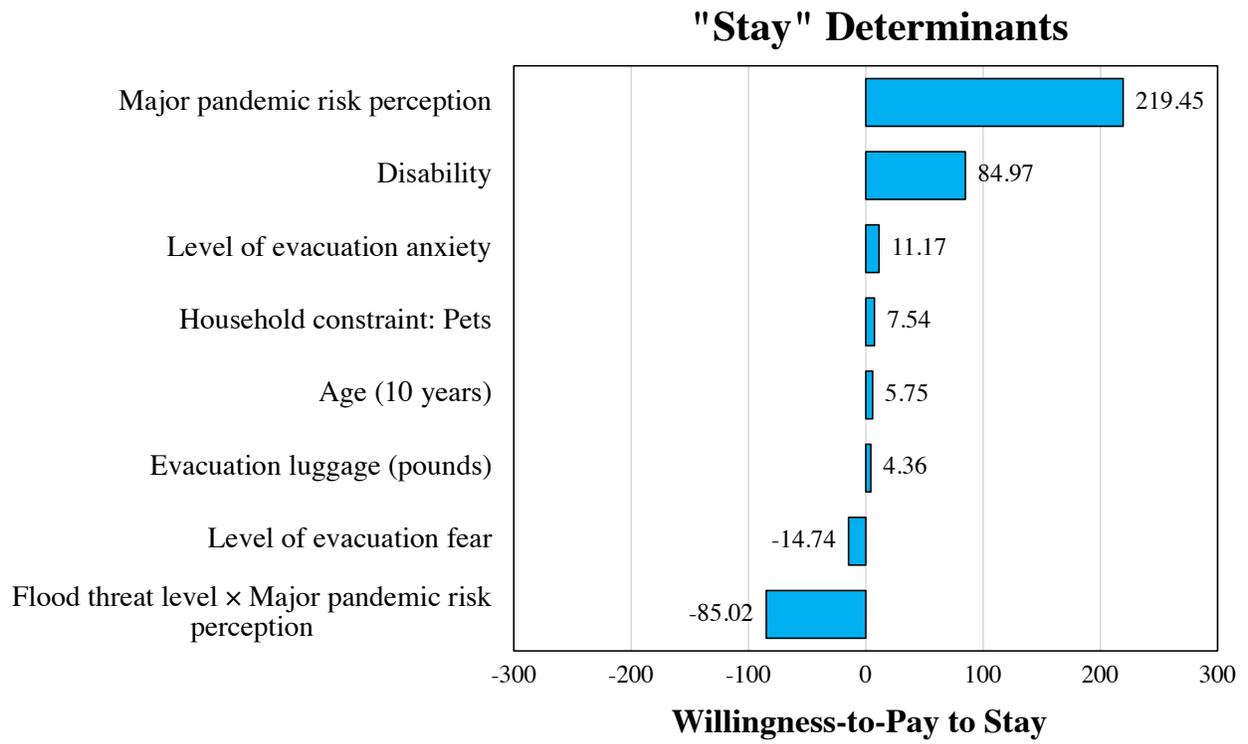